\documentclass[12pt]{article}

\usepackage{amsmath}
\usepackage{graphicx}
\usepackage{algorithm}

\makeatletter
\newcommand\fs@norules{%
  \def\@fs@cfont{\bfseries}\let\@fs@capt\floatc@ruled
  \def\@fs@pre{}%
  \def\@fs@mid{\kern4pt}%
  \def\@fs@post{}%
  \let\@fs@iftopcapt\iftrue}
\makeatother
\floatstyle{norules}
\restylefloat{algorithm}
\usepackage{xcolor}
\usepackage{listings}

\definecolor{codegreen}{rgb}{0.13,0.55,0.13}   
\definecolor{codegray}{rgb}{0.5,0.5,0.5}
\definecolor{codeblue}{rgb}{0.0,0.0,0.8}       
\definecolor{codestring}{rgb}{0.64,0.08,0.08}  
\definecolor{backcolour}{rgb}{0.95,0.95,0.95}

\lstdefinestyle{pythonstyle}{
    language=Python,
    commentstyle=\color{codegreen},
    keywordstyle=\color{codeblue},
    stringstyle=\color{codestring},
}
\title{Twelve Simple Algorithms to Compute Fibonacci Numbers\thanks{The
  algorithms presented here, and their implementations in Python, are also
  available in a public repository~\protect\cite{Da18}. We keep adding new
  algorithms to that repository as we come across them, so it may well
  contain more algorithms than the twelve we present in this paper.}}
\author{
  Ali Dasdan \\
  KD Consulting \\
  Saratoga, CA, USA \\
  alidasdan@gmail.com
}
\date{April 13, 2018\\[2pt]
  {\small (updated on August 30, 2026)}}

\begin{document}
\maketitle

\begin{abstract}
The Fibonacci numbers are a sequence of integers in which every number
after the first two, 0 and 1, is the sum of the two preceding
numbers. These numbers are well known, and the algorithms to compute
them are simple enough that they are often used in introductory
algorithms courses. In this paper, we present twelve such algorithms
together with their time and space complexity analyses. Though very
simple, these algorithms illustrate eleven concepts from the algorithms
field, ranging from top-down vs. bottom-up dynamic programming to
recursion depth, and we say which algorithms illustrate which
concept. We also present the results of a small-scale experimental
comparison of their runtimes on a personal laptop, where the slowest
algorithm takes about four orders of magnitude longer than the
fastest. Finally, we provide a list of homework questions for
students. We hope that this paper can serve as a useful resource for
students learning the basics of algorithms.
\end{abstract}

\section{Introduction}\label{sec:intro}

The {\em Fibonacci numbers} are a sequence $F_n$ of integers in which
every number after the first two, 0 and 1, is the sum of the two
preceding numbers: $0, 1, 1, 2, 3, 5, 8, 13, 21, \ldots$. More formally,
they are defined by the recurrence relation $F_n = F_{n-1} + F_{n-2}$,
$n\geq 2$, with the base values $F_0=0$ and
$F_1=1$~\cite{CrStRi01,GrKnPa94,SoWe07,Wiki18}.

The formal definition of this sequence directly maps to an algorithm
to compute the $n$th Fibonacci number $F_n$. However, there are many
other ways of computing the $n$th Fibonacci number. This paper
presents twelve algorithms in total. Each algorithm takes in $n$ and
returns $F_n$.

Probably due to the simplicity of this sequence, each of the twelve
algorithms is also fairly simple. This allows the use of these
algorithms for teaching some key concepts from the algorithms field,
e.g., see \cite{Fo15} on the use of such algorithms for teaching
dynamic programming. This paper is an attempt in this direction. Among
other things, the algorithmic concepts illustrated by these algorithms
include
\begin{itemize}
\item top-down vs. bottom-up dynamic programming~\cite{CrStRi01,Fo15,Wiki18c},
  illustrated by the top-down \texttt{fib1}, \texttt{fib2} and
  \texttt{fib9} against the bottom-up \texttt{fib3}, \texttt{fib10}
  and \texttt{fib11},
\item dynamic programming with vs. without memoization~\cite{CrStRi01,Fo15,Wiki18c},
  illustrated by \texttt{fib1}, which recomputes every subproblem,
  against \texttt{fib2}, \texttt{fib9}, \texttt{fib10} and
  \texttt{fib11}, which cache them,
\item recursion vs. iteration~\cite{Wiki18d,Wiki18f,Wiki18g},
  illustrated by three pairs that share a formula: \texttt{fib1} and
  \texttt{fib2} against \texttt{fib3}, \texttt{fib7} against
  \texttt{fib8}, and \texttt{fib9} against \texttt{fib10},
\item integer vs. floating-point arithmetic~\cite{Go91,Wiki18h},
  illustrated by \texttt{fib4}, \texttt{fib5} and \texttt{fib12}
  against the nine algorithms that use integer arithmetic only,
\item exact vs. approximate results~\cite{Go91,Wiki18h}, illustrated by
  those same three algorithms, \texttt{fib4} and \texttt{fib5} being
  exact only up to $n=70$ and \texttt{fib12} only up to $n=78$ on our
  experimental setting,
\item exponential-time vs. polynomial-time~\cite{Wiki18e}, illustrated
  by \texttt{fib1} against all the others,
\item constant-time vs. non-constant-time arithmetic~\cite{Wiki18b},
  illustrated by all twelve, in the two columns of
  Table~\ref{tab:cmp},
\item progression from constant to polynomial to exponential time and
  space complexity~\cite{CrStRi01,Wiki18e}, illustrated in time by
  \texttt{fib4} and \texttt{fib5} (constant), \texttt{fib7} through
  \texttt{fib11} (logarithmic), \texttt{fib2}, \texttt{fib3},
  \texttt{fib6} and \texttt{fib12} (linear), and \texttt{fib1}
  (exponential); in space the grouping differs, most sharply for
  \texttt{fib9}, which uses $O(n)$ where \texttt{fib10} and
  \texttt{fib11} use $O(\lg n)$ for the same formula,
\item closed-form vs. recursive formulas~\cite{Wiki18i}, illustrated by
  \texttt{fib4} and \texttt{fib5}, and by the matrix-based
  \texttt{fib6}, \texttt{fib7} and \texttt{fib8}, against the
  recurrence-based \texttt{fib1}, \texttt{fib2}, \texttt{fib3},
  \texttt{fib9}, \texttt{fib10}, \texttt{fib11} and \texttt{fib12},
\item repeated squaring vs. linear iteration for
  exponentiation~\cite{CrStRi01}, illustrated by \texttt{fib6} against
  \texttt{fib7} and \texttt{fib8}, which raise the same matrix to the
  same power, and
\item recursion depth~\cite{Wiki18d,Wiki18f,Wiki18g}, illustrated by
  \texttt{fib1} and \texttt{fib2}, whose depth grows linearly with
  $n$, against \texttt{fib7} and \texttt{fib9}, whose depth grows
  logarithmically.
\end{itemize}
Given the richness of the field, we expect that computing these
numbers will go on yielding further concepts to illustrate.

We present each algorithm as implemented in the Python programming
language (so that they are ready-to-run on a computer) together with
their time and space complexity analyses. We also present a
small-scale experimental comparison of these algorithms. We hope
students with an interest in learning algorithms may find this paper
useful for these reasons. The simplicity of the algorithms should also
help these students to focus on learning the algorithmic concepts
illustrated rather than struggling with understanding the details of
the algorithms themselves.

Since the Fibonacci sequence has been well studied in math, there are
many known relations~\cite{Wiki18}, including the basic recurrence
relation introduced above. Some of the algorithms in this study
directly implement a known recurrence relation on this sequence. Some
others are derived by converting recursion to iteration. In
\cite{Ho88}, even more algorithms together with their detailed
complexity analyses are presented. 

Interestingly, there are also closed-form formulas on the Fibonacci
numbers. The algorithms derived from these formulas are also part of
this study. However, these algorithms produce approximate results
beyond a certain $n$ due to their reliance on the floating-point
arithmetic.

For convenience, we call an algorithm {\em exact} if it returns
exactly $F_n$, and {\em approximate} otherwise.

\section{Preliminaries}\label{sec:prelim}

These are the helper algorithms or functions used by some of the
twelve algorithms. Each of these algorithms are already well known in
the technical literature~\cite{CrStRi01}. They are also simple to
derive.

Note that $m=[[a,b],[c,d]]$ in the Python notation means a 2x2 matrix
\begin{equation}
  m=\left(\begin{array}{cc} a & b\\ c & d \end{array}\right)
\end{equation}
in math notation. Also, each element is marked with its row and column
id as in, e.g., $m[0][1]$ in Python notation means $m_{01}=b$ in math
notation.

Since we will compute $F_n$ for large $n$, the number of bits in $F_n$
will be needed. As we will later see,
\begin{equation}
  F_{n} =
  \left[{\frac{\varphi^n}{\sqrt{5}}}\right]\text{and }\varphi =
  \frac{1+\sqrt{5}}{2} \approx 1.618033
  \label{eq:basic}
\end{equation}
where $\varphi$ is called the {\em golden ratio} and $[\cdot]$ rounds
its argument. Hence, the number of bits in $F_n$ is equal to $\lg F_n
\approx n\lg \varphi \approx 0.7n = \Theta(n)$.

In the sequel, we will use $A(b)$ and $M(b)$ to represent the time
complexity of adding (or subtracting) and multiplying (or dividing)
two $b$-bit numbers, respectively. For fixed precision arguments with
constant width, we will assume that these operations take constant
time, i.e., $A(b)=M(b)=O(1)$; we will refer to this case as {\em
  constant-time arithmetic}. For arbitrary precision arguments,
$A(b)=O(b)$ and $M(b)=O(b^2)$, although improved bounds for each
exist~\cite{Wiki18b}; we will refer to this case as {\em
  non-constant-time arithmetic}. The non-constant case also applies
when we want to express the time complexity in terms of bit
operations, even with fixed point arguments. In this paper, we will
vary $b$ from 32 to $F_{10,000}$, which is around 7,000 bits per
Eq.~\ref{eq:basic}.

What follows next are these helper algorithms, together with their
description, and their time complexity analyses in bit operations.
\begin{itemize}
  
  \item \texttt{is\_odd(n)} and \texttt{is\_even(n)} in
    Alg.~\ref{alg:parity}: Two functions to test the parity of an
    integer $n$. \texttt{is\_odd} masks off all but the lowest bit of
    $n$, which is 1 exactly when $n$ is odd, and \texttt{is\_even}
    negates that. Both take constant time even with non-constant-time
    arithmetic, since only the lowest bit is ever examined. They are
    used by \texttt{negafib} below and by several of the twelve
    algorithms of Section~\ref{sec:algos}.

  \item \texttt{num\_pow\_iter(a,\allowbreak n)} in Alg.~\ref{alg:num_pow_iter}:
    An algorithm to compute $a^n$ for floating-point $a$ and
    non-negative integer $n$ iteratively using {\em repeated
      squaring}, which uses the fact that
    $a^n=(a^{(\frac{n}{2})})^2$. This algorithm iterates over the bits
    of $n$, so it iterates $\lg n$ times. In each iteration, it can
    multiply two $b$-bit numbers at most twice, where $b$ ranges from
    $\lg a$ to $\lg a^{n-1}$ in the worst case, or where each
    iteration takes time from $M(\lg a)$ and $M(\lg a^n)$. The worst
    case happens when the bit string of $n$ is all 1s, i.e., when $n$
    is one less than a power of 2. As such, a trivial worst-case time
    complexity is $O(M(\lg a^n)\lg n)$. However, a more careful
    analysis shaves off the $\lg n$ factor to lead to $O(M(\lg
    a^n))=O(M(n\lg a))$. With constant-time arithmetic, the time
    complexity is $O(\lg n)$. Wherever an algorithm in
    Section~\ref{sec:algos} calls \texttt{math.pow(x,y)} from the
    Python standard library, as \texttt{fib4} and \texttt{fib5} do,
    \texttt{num\_pow\_iter(x,y)} can be used in its place. We state
    that alternative here once rather than repeating it in each
    algorithm.

  \item \texttt{mat\_\allowbreak mul(m1,m2)} in Alg.~\ref{alg:mat_mul}: An
    algorithm to compute the product $m1*m2$ of two 2x2 matrices $m1$
    and $m2$. This algorithm has eight multiplications, so the total
    time complexity is $O(M(max(\lg max(m1), \lg max(m2))))$,
    where $max(m)$ returns the largest element of the matrix $m$.

  \item \texttt{mat\_\allowbreak mul\_\allowbreak opt(m1)} in Alg.~\ref{alg:mat_mul}: An
    algorithm to compute the product $m1*m2$ of two 2x2 matrices $m1$
    and $m2$ when $m2=[[1,1],[1,0]]$, in the Python list
    notation. This algorithm is an optimized version of
    \texttt{mat\_\allowbreak mul(m1,m2)} in Alg.~\ref{alg:mat_mul}. It performs
    two additions and no multiplications, so the total time complexity is
    $O(A(\lg max(m1)))$, and not $O(M(\lg max(m1)))$ as for
    \texttt{mat\_\allowbreak mul(m1,m2)}. Likewise, wherever an
    algorithm calls \texttt{mat\_\allowbreak mul\_\allowbreak opt(m)}, as
    \texttt{fib6} does, the general \texttt{mat\_\allowbreak mul(m,[[1,1],[1,0]])}
    can be used in its place.

  \item \texttt{mat\_pow\_recur(m,n)} in Alg.~\ref{alg:mat_pow_recur}:
    An algorithm to compute $m^n$ of a 2x2 matrix $m$ recursively
    using repeated squaring. Its time complexity analysis is similar
    to that of \texttt{num\_pow\_iter}. As such, the time complexity
    is $O(M(\lg a))$ where $a=max(r)$. With constant-time arithmetic,
    the time complexity is $O(\lg n)$.

  \item \texttt{mat\_pow\_iter(m,n)} in Alg.~\ref{alg:mat_pow_iter}:
    An algorithm to compute $m^n$ of a 2x2 matrix $m$ iteratively
    using repeated squaring. Its time complexity is equal to that of
    its recursive version, i.e., $O(M(\lg a))$ where $a=max(r)$. With
    constant-time arithmetic, the time complexity is $O(\lg n)$.

  \item \texttt{negafib(m,Fn)} in Alg.~\ref{alg:negafib}: An algorithm
    to return the ``negafibonacci'' number corresponding to the $n$th
    Fibonacci number $F_n$. A negafibonacci number is a Fibonacci
    number with a negative index; such numbers are defined as
    $F_{-n}=(-1)^{n+1}F_n$~\cite{Wiki18}.

  \item The function \texttt{round(x)} or \texttt{[x]} rounds its
    floating-point argument to the closest integer. It maps to the
    \texttt{math.round(x)} function from the standard math library of
    Python. 
\end{itemize}

\begin{algorithm}[htbp]
  \caption{\texttt{is\_odd($n$)} and \texttt{is\_even($n$)}: Two
    algorithms to test whether the integer $n$ is odd or even.}
  \label{alg:parity}
\begin{lstlisting}[style=pythonstyle]
def is_odd(n:int) -> int:
    return n & 1

def is_even(n:int) -> bool:
    return not is_odd(n)
\end{lstlisting}
\end{algorithm}

\begin{algorithm}[htbp]
  \caption{\texttt{num\_pow\_iter($a$,$n$)}: An algorithm to compute $a^n$
    for floating-point $a$ and non-negative integer $n$ iteratively
    using repeated squaring.}
  \label{alg:num_pow_iter}
\begin{lstlisting}[style=pythonstyle]
def num_pow_iter(a:int, n:int) -> int:
    n_bits = bin(n)[2:]
    r = 1
    for b in n_bits:
        r = r * r
        if b == '1':
            r *= a
    return r
\end{lstlisting}
\end{algorithm}

\begin{algorithm}[htbp]
  \caption{\texttt{mat\_mul($m_1$,$m_2$)}: An algorithm to compute the
    product $m_1 m_2$ of two 2x2 matrices $m_1$ and
    $m_2$. \texttt{mat\_mul\_opt($m_1$)} is the optimized version for
    the case $m_2=[[1,1],[1,0]]$.}
  \label{alg:mat_mul}
\begin{lstlisting}[style=pythonstyle]
def mat_mul(m1:Matrix, m2:Matrix) -> Matrix:
    m = [[0, 0], [0, 0]]
    m[0][0] = m1[0][0] * m2[0][0] + m1[0][1] * m2[1][0]
    m[0][1] = m1[0][0] * m2[0][1] + m1[0][1] * m2[1][1]
    m[1][0] = m1[1][0] * m2[0][0] + m1[1][1] * m2[1][0]
    m[1][1] = m1[1][0] * m2[0][1] + m1[1][1] * m2[1][1]
    return m

def mat_mul_opt(m1:Matrix) -> Matrix:
    m = [[0, 0], [0, 0]]
    m[0][0] = m1[0][0] + m1[0][1]
    m[0][1] = m1[0][0]
    m[1][0] = m1[1][0] + m1[1][1]
    m[1][1] = m1[1][0]
    return m
\end{lstlisting}
\end{algorithm}

\begin{algorithm}[htbp]
  \caption{\texttt{mat\_pow\_recur($m$,$n$)}: An algorithm to compute
    $m^n$ of a 2x2 matrix $m$ for non-negative integer $n$ recursively
    using repeated squaring. \texttt{mat\_pow\_opt\_recur($n$)} is the
    optimized version for the case $m=[[1,1],[1,0]]$.}
  \label{alg:mat_pow_recur}
\begin{lstlisting}[style=pythonstyle]
def mat_pow_recur(m:Matrix, n:int) -> Matrix:
    r = [[1, 0], [0, 1]] # identity
    if n > 1:
        r = mat_pow_recur(m, n >> 1)
        r = mat_mul(r, r)
    if is_odd(n):
        r = mat_mul(r, m)
    return r

def mat_pow_opt_recur(n:int) -> Matrix:
    r = [[1, 0], [0, 1]] # identity
    if n > 1:
        r = mat_pow_opt_recur(n >> 1)
        r = mat_mul(r, r)
    if is_odd(n):
        r = mat_mul_opt(r)
    return r
\end{lstlisting}
\end{algorithm}

\begin{algorithm}[htbp]
  \caption{\texttt{mat\_pow\_iter($m$,$n$)}: An algorithm to compute $m^n$
    of a 2x2 matrix $m$ for non-negative integer $n$ iteratively using
    repeated squaring. \texttt{mat\_pow\_opt\_iter($n$)} is the
    optimized version for the case $m=[[1,1],[1,0]]$.}
  \label{alg:mat_pow_iter}
\begin{lstlisting}[style=pythonstyle]
def mat_pow_iter(m:Matrix, n:int) -> Matrix:
    n_bits = bin(n)[2:]
    r = [[1, 0], [0, 1]] # identity
    for b in n_bits:
        r = mat_mul(r, r)
        if b == '1':
            r = mat_mul(r, m)
    return r

def mat_pow_opt_iter(n:int) -> Matrix:
    n_bits = bin(n)[2:]
    r = [[1, 0], [0, 1]] # identity
    for b in n_bits:
        r = mat_mul(r, r)
        if b == '1':
            r = mat_mul_opt(r)
    return r
\end{lstlisting}
\end{algorithm}

\begin{algorithm}[htbp]
  \caption{\texttt{negafib($n$,$F_n$)}: An algorithm to compute the
    negafibonacci number corresponding to the $n$th Fibonacci number
    $F_n$, using \texttt{is\_odd} from Alg.~\ref{alg:parity}.}
  \label{alg:negafib}
\begin{lstlisting}[style=pythonstyle]
def negafib(n:int, Fn:int) -> int:
    if is_odd(n + 1):
        return -Fn
    return Fn
\end{lstlisting}
\end{algorithm}

\section{The Twelve Algorithms}\label{sec:algos}

We now present each algorithm, together with a short explanation on
how it works, its time and space complexity analyses, and some of the
concepts from the algorithms field it illustrates. Each algorithm
takes $n$ as the input and returns the $n$th Fibonacci number
$F_n$. Note that $n$ can also be negative, in which case the returned
numbers are called ``negafibonacci'' numbers, defined as
$F_{-n}=(-1)^{n+1}F_n$~\cite{Wiki18b}.

Some of the algorithms use a data structure $F$ to cache pre-computed
numbers such that $F[n]=F_n$. For some algorithms $F$ is an array (a
mutable list in Python) whereas for some others it is a hash table (or
dictionary in Python). In Python, lists and dictionaries are accessed
the same way, so the type of the data structure can be found out by
noting whether or not the indices accessed are consecutive or not.

Each algorithm is also structured such that the base cases
for $n=0$ to $n=1$ (or $n=2$ in some cases) are taken care of before
the main part is run. Also note that some of the algorithms in this
section rely on one or more algorithms from the preliminaries section.

The listings below are the Python 3 implementations from the
repository~\cite{Da18}. There, each algorithm lives in its own module
\texttt{ad\_fibN.py} and its entry point is named simply
\texttt{fib}; we write \texttt{fibN} in the listings so that the
twelve can be told apart.

\begin{algorithm}[htbp]
  \caption{\texttt{fib1($n$)}: An algorithm to compute the $n$th
    Fibonacci number $F_n$ recursively using dynamic programming. This
    algorithm uses probably the most well-known recurrence relation
    defining Fibonacci numbers.}
  \label{alg:fib1}
\begin{lstlisting}[style=pythonstyle]
def fib_recur(n:int) -> int:
    if n == 0:
        return 0
    if n == 1:
        return 1
    return fib_recur(n - 1) + fib_recur(n - 2)

def fib1(n:int) -> int:
    n0, n = n, abs(n)
    r = fib_recur(n)
    if n0 < 0:
        return negafib(n, r)
    return r
\end{lstlisting}
\end{algorithm}

\subsection{Algorithm fib1: Top-down Dynamic Programming}

\texttt{fib1} is derived directly from the recursive definition of the
Fibonacci sequence: $F_n = F_{n-1} + F_{n-2}$, $n\geq 2$, with the
base values $F_0=0$ and $F_1=1$.

Its time complexity $T(n)$ with constant-time arithmetic can be
expressed as $T(n)=T_{n-1}+T_{n-2}+1$, $n\geq 2$, with the base values
$T(0)=1$ and $T(1)=2$. The solution of this linear non-homogeneous
recurrence relation implies that the time complexity is equal to
$T(n)=O(F_{n})=O(\varphi^n)$, which is exponential in $n$.

Its time complexity with non-constant-time arithmetic leads to
$T(n)=T_{n-1}+T_{n-2}+A(\lg F_{n-1})$, where the last term signifying
the cost of the addition run in $O(n)$ time. The solution of this
linear non-homogeneous recurrence relation implies that the time
complexity is also equal to $T(n)=O(F_{n})=O(\varphi^n)$, which is
also exponential in $n$.

The space complexity, in contrast, is not exponential. The recursion
tree has $O(\varphi^n)$ nodes, but only the nodes on the path from the
root down to the current node are live at any moment. Hence the
recursion depth, and with it the space complexity, is $O(n)$ with
constant-time arithmetic and $O(n^2)$ bits with non-constant-time
arithmetic~\cite{Wiki18d,Wiki18f}.

Regarding algorithmic concepts, \texttt{fib1} illustrates recursion,
top-down dynamic programming, and exponential complexity.

\begin{algorithm}[htbp]
  \caption{\texttt{fib2($n$)}: An algorithm to compute the $n$th
    Fibonacci number $F_n$ recursively using dynamic programming with
    memoization (i.e., caching of pre-computed results).}
  \label{alg:fib2}
\begin{lstlisting}[style=pythonstyle]
def fib_recur(n:int, F:List[int]) -> int:
    if F[n] is None:
        if n == 0:
            F[n] = 0
        elif n == 1:
            F[n] = 1
        else:
            F[n] = fib_recur(n - 1, F) + fib_recur(n - 2, F)
    return F[n]

def fib2(n:int) -> int:
    n0, n = n, abs(n)
    r = fib_recur(n, [None] * (n + 1))
    if n0 < 0:
        return negafib(n, r)
    return r
\end{lstlisting}
\end{algorithm}

\subsection{Algorithm fib2: Top-down Dynamic Programming with Memoization}

\texttt{fib2} is equivalent to \texttt{fib1} except for the addition of
memoization. Memoization allows the caching of the already
computed Fibonacci numbers so that \texttt{fib2} does not have to
revisit already visited parts of the call tree.

Memoization reduces the time and space complexities drastically; it
leads to at most $n$ additions. With constant-time arithmetic, the time
complexity is $O(n)$. The space complexity is also linear.

With non-constant time arithmetic, the additions range in time
complexity from $A(\lg F_2)$ to $A(\lg F_{n-1})$. The sum of these
additions leads to the time complexity of $O(n^2)$ in bit
operations. The space complexity is also quadratic in the number of
bits.

Regarding algorithmic concepts, \texttt{fib2} illustrates recursion,
top-down dynamic programming, and memoization.

\begin{algorithm}[htbp]
  \caption{\texttt{fib3($n$)}: An algorithm to compute the $n$th
    Fibonacci number $F_n$ iteratively in constant storage.}
  \label{alg:fib3}
\begin{lstlisting}[style=pythonstyle]
def fib3(n:int) -> int:
    n0, n = n, abs(n)
    if n == 0:
        r = 0
    elif n == 1:
        r = 1
    else:
        f2, f1 = 1, 0
        for _ in range(2, n + 1):
            f2, f1 = f2 + f1, f2
        r = f2
    if n0 < 0:
        return negafib(n, r)
    return r
\end{lstlisting}
\end{algorithm}

\subsection{Algorithm fib3: Iteration with Constant Storage}

\texttt{fib3} uses the fact that each Fibonacci number depends only on
the preceding two numbers in the Fibonacci sequence. This fact turns
an algorithm designed top down, namely, \texttt{fib2}, to one designed
bottom up. This fact also reduces the space usage to constant, just a
few variables.

The time complexity is exactly the same as that of \texttt{fib2}. The
space complexity is $O(1)$ with constant-time arithmetic and $O(n)$
with non-constant-time arithmetic.

Regarding the algorithmic concepts, \texttt{fib3} illustrates
iteration, recursion to iteration conversion, bottom-up dynamic
programming, and constant space.

\begin{algorithm}[htbp]
  \caption{\texttt{fib4($n$)}: An algorithm to compute the $n$th
    Fibonacci number $F_n$ using the closed-form formula with the
    golden ratio. It is exact up to $n=70$ on our
    experimental setting, and raises beyond $n=1474$, where
    $\varphi^n$ leaves the IEEE 754 double precision range.}
  \label{alg:fib4}
\begin{lstlisting}[style=pythonstyle]
def fib4(n:int) -> int:
    n0, n = n, abs(n)
    if n == 0:
        r = 0
    elif n == 1:
        r = 1
    else:
        sqrt_5 = math.sqrt(5)
        phi = float(1 + sqrt_5) / 2
        phi_n = math.pow(phi, n)
        psi_n = float(1) / phi_n
        if is_odd(n):
            psi_n = -psi_n
        r = round((phi_n - psi_n) / sqrt_5)
        r = int(r)
    if n0 < 0:
        return negafib(n, r)
    return r
\end{lstlisting}
\end{algorithm}

\subsection{Algorithm fib4: Closed-form Formula with the Golden Ratio}

\texttt{fib4} uses the fact that the $n$th Fibonacci number has the
following closed-form formula~\cite{SoWe07,Wiki18}:
\begin{equation}
  \label{eq:closed1}
  F_{n} = \left[\frac{\varphi^n - \psi^n}{\varphi - \psi}\right]
  =\left[\frac{\varphi^n - \psi^n}{\sqrt{5}}\right]
\end{equation}
where
\begin{equation}
  \varphi = \frac{1+\sqrt{5}}{2} \approx 1.618033
\end{equation}
is the golden ratio,
\begin{equation}
  \psi = \frac{1-\sqrt{5}}{2} = 1-\psi=-\frac{1}{\varphi}\approx -0.618033
\end{equation}
is the negative of its conjugate.

Note that to compute $\varphi^n$, this algorithm uses the standard
math library function \texttt{math.pow(phi,n)}; as noted in
Section~\ref{sec:prelim}, \texttt{num\_pow\_iter(phi,n)} can be used in
its place.

This algorithm performs all its functions in floating-point
arithmetic; the math functions in the algorithm map to machine
instructions directly. Thus, this algorithm runs in constant time and
space (also see \cite{Ho88} for an argument on non-constant time,
assuming large $n$).

For $n>70$, this algorithm starts returning approximate results. For
$n>1474$, this algorithm starts erroring out as $F_n$ is too large to
even fit in a double precision floating point number width. These two
limits have different origins. The first depends on the 53-bit
significand of the double precision format and on the accuracy of the
library's \texttt{pow} and \texttt{sqrt}, which IEEE 754 does not
require to be correctly rounded; it may therefore change with the
programming language, the math library, and the computer used, and so
may the corresponding limits of \texttt{fib5} and \texttt{fib12}. The
second follows from the exponent range alone: the largest finite
double is about $1.8\times 10^{308}$, and since $\lfloor\ln(1.8\times
10^{308})/\ln\varphi\rfloor=1474$, the value $\varphi^{n}$ ceases to be
representable at $n=1475$. That limit is the same wherever the IEEE
754 double precision format is used. The general observations still
hold.

Regarding the algorithmic concepts, \texttt{fib4} illustrates the
closed-form formula vs. iteration, integer vs. floating-point
computation, and the exact vs. approximate result (or computation).

\begin{algorithm}[htbp]
  \caption{\texttt{fib5($n$)}: An algorithm to compute the $n$th
    Fibonacci number $F_n$ using the closed-form formula with the
    golden ratio and rounding. It is exact up to $n=70$ on our
    experimental setting, and raises beyond $n=1474$, where
    $\varphi^n$ leaves the IEEE 754 double precision range.}
  \label{alg:fib5}
\begin{lstlisting}[style=pythonstyle]
def fib5(n:int) -> int:
    n0, n = n, abs(n)
    if n == 0:
        r = 0
    elif n == 1:
        r = 1
    else:
        sqrt_5 = math.sqrt(5)
        phi = float(1 + sqrt_5) / 2
        phi_n = math.pow(phi, n)
        r = round(phi_n / sqrt_5)
        r = int(r)
    if n0 < 0:
        return negafib(n, r)
    return r
\end{lstlisting}
\end{algorithm}

\subsection{Algorithm fib5: Closed-form Formula with the Golden Ratio and
  Rounding}

\texttt{fib5} relies on the same closed-form formula used by
\texttt{fib4}. However, it also uses the fact that $\psi$ is less than
1, meaning its $n$th power for large $n$ approaches
zero~\cite{SoWe07,Wiki18}. This means Eq.~\ref{eq:closed1} reduces to
\begin{equation}
  \label{eq:closed2}
  F_{n} = \left[{\frac{\varphi^n}{\sqrt{5}}}\right]
\end{equation}
where $\varphi$ is the golden ratio.

For the same reasons as in \texttt{fib4}, this algorithm also runs in
constant time and space.

The approximate behaviour of this algorithm is the same as that of
\texttt{fib4}.

Regarding the algorithmic concepts, \texttt{fib5} illustrates the
concept of the optimization of a closed-form formula for speed-up in
addition to the algorithmic concepts illustrated by \texttt{fib4}.

\begin{algorithm}[htbp]
  \caption{\texttt{fib6($n$)}: An algorithm to compute the $n$th
    Fibonacci number $F_n$ using the power of a certain 2x2 matrix,
    computed iteratively by repeated multiplication.}
  \label{alg:fib6}
\begin{lstlisting}[style=pythonstyle]
def fib6(n:int) -> int:
    n0, n = n, abs(n)
    if n == 0:
        r = 0
    else:
        m = [[1, 0], [0, 1]]
        for _ in range(1, n):
            m = mat_mul_opt(m)
        r = m[0][0]
    if n0 < 0:
        return negafib(n, r)
    return r
\end{lstlisting}
\end{algorithm}

\subsection{Algorithm fib6: The Power of a Certain 2x2 Matrix via Iteration}

\texttt{fib6} uses the power of a certain 2x2 matrix for the Fibonacci
sequence~\cite{SoWe07,Wiki18}:
\begin{equation}
  \left(\begin{array}{cc} 1 & 1\\ 1 & 0 \end{array}\right)^{n-1}
  =
  \left(\begin{array}{cc} F_{n} & F_{n-1}\\ F_{n-1} & F_{n-2} \end{array}\right)
\end{equation}
where $n\geq 2$. Then, $F_n$ is the largest element of the resulting
2x2 matrix.

The time complexity analysis here is similar to that of
\texttt{fib3}. With constant-time arithmetic, the time complexity is
$O(n)$. Unlike \texttt{fib2}, this algorithm keeps only a single 2x2
matrix rather than a table of all the previous values, so its space
complexity is $O(1)$, as in \texttt{fib3}.

With non-constant time arithmetic, the additions range in time
complexity from $A(\lg F_1)$ to $A(\lg F_{n})$. The sum of these
additions leads to the time complexity of $O(n^2)$ in bit
operations. The space complexity is $O(n)$ in the number of bits,
since each of the four matrix entries holds $O(n)$ bits.

Regarding the algorithmic concepts, \texttt{fib6} illustrates (simple)
matrix algebra and iteration over closed-form equations.

\begin{algorithm}[htbp]
  \caption{\texttt{fib7($n$)}: An algorithm to compute the $n$th
    Fibonacci number $F_n$ using the power of a certain 2x2 matrix,
    computed recursively by repeated squaring.}
  \label{alg:fib7}
\begin{lstlisting}[style=pythonstyle]
def fib7(n:int) -> int:
    n0, n = n, abs(n)
    if n == 0:
        r = 0
    else:
        m = mat_pow_opt_recur(n - 1)
        r = m[0][0]
    if n0 < 0:
        return negafib(n, r)
    return r
\end{lstlisting}
\end{algorithm}

\begin{algorithm}[htbp]
  \caption{\texttt{fib8($n$)}: An algorithm to compute the $n$th
    Fibonacci number $F_n$ using the power of a certain 2x2 matrix,
    computed iteratively by repeated squaring.}
  \label{alg:fib8}
\begin{lstlisting}[style=pythonstyle]
def fib8(n:int) -> int:
    n0, n = n, abs(n)
    if n == 0:
        r = 0
    else:
        m = mat_pow_opt_iter(n - 1)
        r = m[0][0]
    if n0 < 0:
        return negafib(n, r)
    return r
\end{lstlisting}
\end{algorithm}

\subsection{Algorithms fib7 and fib8: The Power of a Certain 2x2
  Matrix via Repeated Squaring}

\texttt{fib7} and \texttt{fib8} use the same equations used in
\texttt{fib6} but while \texttt{fib6} uses iteration for
exponentiation, \texttt{fib7} and \texttt{fib8} uses repeated squaring
for speed-up. Moreover, \texttt{fib7} uses a recursive version of
repeated squaring while \texttt{fib8} uses an iterative version of it.

Repeated squaring reduces time from linear to logarithmic. Hence, with
constant time arithmetic, the time complexity is $O(\lg n)$. The space
complexity is also logarithmic in $n$.

With non-constant time arithmetic, the cost is dominated by the
multiplications rather than by the additions. Each squaring step
multiplies two $b$-bit numbers, where $b$ grows to $\lg
F_{n}=\Theta(n)$ at the last step. Summing over the $\lg n$ steps gives
a geometric series whose last term dominates, so the time complexity
is $O(M(n))$ in bit operations, which is $O(n^2)$ under the schoolbook
bound $M(b)=O(b^2)$ assumed in Section~\ref{sec:prelim}. This agrees
with the analyses of \texttt{mat\_pow\_recur} and
\texttt{mat\_pow\_iter} given there. The space complexity is linear in
the number of bits.

Regarding the algorithmic concepts, \texttt{fib7} and \texttt{fib8}
illustrate (simple) matrix algebra, and repeated squaring over
closed-form equations. In addition, \texttt{fib7} illustrates
recursion while \texttt{fib8} illustrates iteration to perform
repeated squaring over a matrix.

\begin{algorithm}[htbp]
  \caption{\texttt{fib9($n$)}: An algorithm to compute the $n$th
    Fibonacci number $F_n$ recursively using a certain recursive
    formula, with memoization.}
  \label{alg:fib9}
\begin{lstlisting}[style=pythonstyle]
def fib_recur(n:int, F:List[int]) -> int:
    if F[n] is None:
        if n <= 0:
            F[n] = 0
        elif n == 1:
            F[n] = 1
        elif n == 2:
            F[n] = 1
        else:
            k = n >> 1
            f1 = fib_recur(k, F)
            f2 = fib_recur(k + 1, F)
            if is_even(n):
                F[n] = 2 * f1 * f2 - f1 * f1
            else:
                F[n] = f2 * f2 + f1 * f1
    return F[n]

def fib9(n:int) -> int:
    n0, n = n, abs(n)
    r = fib_recur(n, [None] * (n + 1))
    if n0 < 0:
        return negafib(n, r)
    return r
\end{lstlisting}
\end{algorithm}

\begin{algorithm}[htbp]
  \caption{\texttt{fib10($n$)}: An algorithm to compute the $n$th
    Fibonacci number $F_n$ iteratively using the same recursive
    formula as \texttt{fib9}. $F$ is a hash table and $B$ holds the
    base values $B[0]=0$, $B[1]=1$, and $B[2]=1$.}
  \label{alg:fib10}
\begin{lstlisting}[style=pythonstyle]
BASE = {0: 0, 1: 1, 2: 1}

def fib10(n:int) -> int:
    n0, n = n, abs(n)

    # find indexes that need F values. None marks an index whose value
    # is not known yet.
    F = {n: None}
    qinx = [n]   # queue of indexes
    while qinx:
        k = qinx.pop() >> 1
        for j in (k, k + 1):
            if j not in F:
                F[j] = None
                qinx.append(j)

    # set base values
    F.update(BASE)

    # fill the indexes that need values. the increasing index order
    # guarantees that F[k >> 1] and F[(k >> 1) + 1] are already known.
    for k in sorted(F.keys()):
        if k in BASE:
            continue
        k2 = k >> 1
        f1, f2 = F[k2], F[k2 + 1]
        if is_even(k):
            F[k] = 2 * f2 * f1 - f1 * f1
        else:
            F[k] = f2 * f2 + f1 * f1

    r = F[n]
    if n0 < 0:
        return negafib(n, r)
    return r
\end{lstlisting}
\end{algorithm}

\subsection{Algorithms fib9 and fib10: A Certain Recursive Formula}

Both \texttt{fib9} and \texttt{fib10} use the following formulas for
the Fibonacci sequence~\cite{Wiki18}.
\begin{equation}
  F_{2n+1}=F_{n+1}^2+F_{n}^2\mbox{ and }
  F_{2n}=2F_{n+1}F_{n}-F_{n}^2
\end{equation}
where $n\geq 2$, with the base values $F_2=1$, $F_1=1$, and $F_0=0$. Note
that memoization is used for speed-up in such a way that only the
cells needed for the final result are filled in the memoization table
$F$. In \texttt{fib9}, recursion takes care of identifying such
cells. In \texttt{fib10}, a cell marking phase, using a separate queue
of indexes, is used for such identification; then the values for
these cells, starting from the base values, are computed in a
bottom-up fashion.

These algorithms behave like repeated squaring in terms of time
complexity. Hence, with constant time arithmetic, the time complexity
is $O(\lg n)$. Only $O(\lg n)$ of the values $F[k]$ are ever needed,
but the two algorithms store them differently: \texttt{fib10} keeps
exactly those in a hash table, so its space complexity is $O(\lg n)$,
whereas \texttt{fib9} allocates an array of $n+1$ cells and fills only
$O(\lg n)$ of them, so its space complexity is $O(n)$. This is the
array vs. hash table distinction noted at the start of
Section~\ref{sec:algos}.

With non-constant time arithmetic, the cost is again dominated by the
multiplications in the doubling formulas rather than by the additions,
since the last step multiplies $\Theta(n)$-bit numbers. As in
\texttt{fib7} and \texttt{fib8}, summing over the $\lg n$ steps gives
a time complexity of $O(M(n))$ in bit operations, which is $O(n^2)$
under the schoolbook bound. The values stored take $O(n)$ bits in both
algorithms.

Regarding the algorithm concepts, these algorithms illustrate
recursion vs. iteration, top-down vs. bottom-up processing and/or
dynamic programming, implementation of a recursive relation, and
careful use of a queue data structure to eliminate unnecessary work in
the bottom-up processing.

\begin{algorithm}[htbp]
  \caption{\texttt{fib11($n$)}: An algorithm to compute the $n$th
    Fibonacci number $F_n$ iteratively using yet another recursive
    formula. $F$ and $B$ are as in \texttt{fib10}.}
  \label{alg:fib11}
\begin{lstlisting}[style=pythonstyle]
BASE = {0: 0, 1: 1, 2: 1}

def fib11(n:int) -> int:
    n0, n = n, abs(n)

    # find indexes that need F values. None marks an index whose value
    # is not known yet. negative indexes are never needed: they arise
    # only for k <= 1, which the base values already cover.
    F = {n: None}
    qinx = [n]  # queue of indexes
    while qinx:
        k = qinx.pop() >> 1
        for j in (k - 1, k, k + 1):
            if j >= 0 and j not in F:
                F[j] = None
                qinx.append(j)

    # set base values
    F.update(BASE)

    # fill the indexes that need values. the increasing index order
    # guarantees that F[(k >> 1) - 1], F[k >> 1] and F[(k >> 1) + 1]
    # are already known.
    for k in sorted(F.keys()):
        if k in BASE:
            continue
        k2 = k >> 1
        f1, f2, f3 = F[k2 - 1], F[k2], F[k2 + 1]
        if is_even(k):
            F[k] = f3 * f3 - f1 * f1
        else:
            F[k] = f3 * f3 + f2 * f2

    r = F[n]
    if n0 < 0:
        return negafib(n, r)
    return r
\end{lstlisting}
\end{algorithm}

\subsection{Algorithm fib11: Yet Another Recursive Formula}
\texttt{fib11} uses the following formulas for the Fibonacci
sequence~\cite{Wiki18}.
\begin{equation}
  F_{2n+1}=F_{n+1}^2+F_{n}^2\mbox{ and }
  F_{2n}=F_{n+1}^2-F_{n-1}^2
\end{equation}
where $n\geq 2$. The case for $n=2$ needs to be handled as a base case of
recursion to prevent an infinite loop. Note that memoization is also
used for speed-up.

The time and space complexity analyses are as in \texttt{fib10},
which uses the same hash table structure.

Regarding the algorithmic concepts, this algorithm is again similar to
\texttt{fib9} and \texttt{fib10}.

\begin{algorithm}[htbp]
  \caption{\texttt{fib12($n$)}: An algorithm to compute the $n$th
    Fibonacci number $F_n$ iteratively using a certain recursive
    formula built on the golden ratio.}
  \label{alg:fib12}
\begin{lstlisting}[style=pythonstyle]
def fib12(n:int) -> int:
    n0, n = n, abs(n)
    sqrt_5 = math.sqrt(5)
    phi = float(1 + sqrt_5) / 2
    if n == 0:
        r = 0
    elif n == 1:
        r = 1
    elif n == 2:
        r = 1
    else:
        r = 1
        for _ in range(3, n + 1):
            r = round(phi * r)
        r = int(r)
    if n0 < 0:
        return negafib(n, r)
    return r
\end{lstlisting}
\end{algorithm}

\subsection{Algorithm fib12: Yet Another but Simpler Recursive Formula}
\texttt{fib12} uses the following formula for the Fibonacci
sequence
\begin{equation}
  F_{n}=\left[\varphi F_{n-1}\right]
\end{equation}
where $n \geq 3$. It is formula 64 in Vajda~\cite{Va89} and formula 73
in Dunlap~\cite{Du97}, both stated in
the equivalent reindexed form $F_{n+1}=\left[\varphi F_{n}\right]$ for
$n\geq 2$, and it is listed with those attributions in Knott's
collection of Fibonacci and golden ratio formulae~\cite{Kno}.

The identity follows from Eq.~\ref{eq:closed1} in one step. Substituting
$F_{n}=(\varphi^n-\psi^n)/\sqrt{5}$ and using $\psi-\varphi=-\sqrt{5}$,
\begin{equation}
  \varphi F_{n-1}-F_{n}
  = \frac{\varphi^{n}-\varphi\psi^{n-1}-\varphi^{n}+\psi^{n}}{\sqrt{5}}
  = \frac{\psi^{n-1}(\psi-\varphi)}{\sqrt{5}}
  = -\psi^{n-1}.
  \label{eq:phistep}
\end{equation}
So $\varphi F_{n-1}$ misses $F_{n}$ by exactly $\psi^{n-1}$. Since
$|\psi|\approx 0.618$, we get $|\psi^{n-1}|<\frac{1}{2}$ precisely when
$n\geq 3$, which is where the rounding starts to be correct: at $n=2$,
$\varphi F_{1}\approx 1.618$ rounds to 2 rather than to $F_{2}=1$. Note
that this bound is on the formula itself, in exact arithmetic; the
value of $n$ beyond which \texttt{fib12} returns approximate results is
set instead by the floating-point precision it uses.

Note that although the golden ratio is only the limit value of the
ratio of the consecutive Fibonacci numbers,
Eq.~\ref{eq:phistep} shows that the formula is not merely asymptotic:
the rounding is correct for every $n\geq 3$, not just for large $n$.
Our implementation returns exact results from $n=3$ to $n=78$ and
approximate ones beyond that, like the algorithms \texttt{fib4} and
\texttt{fib5}, due to the use of floating-point arithmetic.

The time and space complexity analyses are as in \texttt{fib3}. It is
easy to implement a version of this algorithm where both recursion and
memoization are used.

Regarding the algorithmic concepts, this algorithm illustrates the
iterative version of a recursive formula.

\section{Results}\label{sec:results}

We now present the results of a small-scale experimental analysis done
on a high-end laptop computer (model: MacBook Pro, operating system:
macOS Sierra 10.12.6, CPU: 2.5 GHz Intel Core i7, main memory: 16GB
1600 MHz DDR3).

For each algorithm, we measure its runtime in seconds (using the
\texttt{perf\_\allowbreak counter()} method from the \texttt{time} module in the
Python standard library). We ran each algorithm 10,000 times and collected
the runtimes. The reported runtimes are the averages over these
repetitions. We also computed the standard deviation over these
runtimes. We use the standard deviation to report ``the coefficient of
variability'' (CV) (the ratio of the standard deviation to the average
runtime of an algorithm over 10,000 repetitions), which gives an idea
on the variability of the runtimes.

We report the results in four groups, moving the focus towards the
fastest algorithms as $n$ gets larger:
\begin{enumerate}
  
\item The case $0\leq n\leq 30$ is small enough to run all algorithms,
  including the slowest algorithm \texttt{fib1}.

\item The case $0\leq n\leq 70$ excludes the slowest algorithm
  \texttt{fib1}. In this case all algorithms are exact. On our
  experimental setting, \texttt{fib4} and \texttt{fib5} start returning
  approximate results after $n=70$, and \texttt{fib12} after $n=78$.

\item The case $0\leq n\leq 900$ excludes the approximation
  algorithms, i.e., \texttt{fib4}, \texttt{fib5}, and
  \texttt{fib12}. The upper bound $900$ is also roughly the upper
  bound beyond which the recursive algorithms start exceeding their
  maximum recursion depth limit and error out.

\item The case $0\leq n\leq 10,000$ focuses on the fastest algorithms
  only, excluding the slow algorithms, the recursive algorithms, and
  the approximate algorithms.

\end{enumerate}

Each plot in the sequel has two axes: The x-axis is $n$, as in the
index of $F_n$; and the y-axis is the average runtime in seconds over
10,000 repetitions.

To rank the algorithms in runtime, we use the sum of all the runtimes
across all the $n$ range. These rankings should be taken as
directionally correct as the variability in runtimes, especially for
small $n$, makes it difficult to assert a definite ranking.

\begin{table}[]
\centering
\caption{Algorithms}
\label{tab:cmp}
\begin{tabular}{lcc}
  \hline
  & Runtime in & Runtime in \\
  Algorithm & constant-time ops & bit ops \\
\hline\hline
\texttt{fib1} &  $O(\varphi^n)$ &  $O(\varphi^n)$ \\
\hline
 \texttt{fib2} &  $O(n)$ & $O(n^2)$ \\
\hline
 \texttt{fib3} &  $O(n)$ & $O(n^2)$ \\
\hline
 \texttt{fib4} &  $O(1)$ & $O(1)$ \\
\hline
 \texttt{fib5} &  $O(1)$ & $O(1)$ \\
\hline
 \texttt{fib6} &  $O(n)$ & $O(n^2)$ \\
\hline
 \texttt{fib7} &  $O(\lg n)$ & $O(M(n))=O(n^2)$ \\
\hline
 \texttt{fib8} &  $O(\lg n)$ & $O(M(n))=O(n^2)$ \\
\hline
 \texttt{fib9} &  $O(\lg n)$ & $O(M(n))=O(n^2)$ \\
\hline
 \texttt{fib10} &  $O(\lg n)$ & $O(M(n))=O(n^2)$ \\
\hline
 \texttt{fib11} &  $O(\lg n)$ & $O(M(n))=O(n^2)$ \\
\hline
 \texttt{fib12} &  $O(n)$ & $O(n^2)$ \\
\hline
\end{tabular}

\vspace{6pt}
\begin{minipage}{0.92\textwidth}
{\small The bit-operation column uses the schoolbook bounds $A(b)=O(b)$
and $M(b)=O(b^2)$ from Section~\ref{sec:prelim}. A subquadratic
multiplication improves every $O(M(n))$ entry: the Python
implementations use Karatsuba multiplication, for which
$M(b)=O(b^{1.585})$. The $O(1)$ entries for \texttt{fib4} and
\texttt{fib5} count the floating-point arithmetic only; any algorithm
that returns $F_n$ as an exact integer must write $\Theta(n)$ bits, so
$\Omega(n)$ is a lower bound on the output alone.}
\end{minipage}
\end{table}

\begin{figure}
  \centering
  \includegraphics[width=0.7\textwidth]{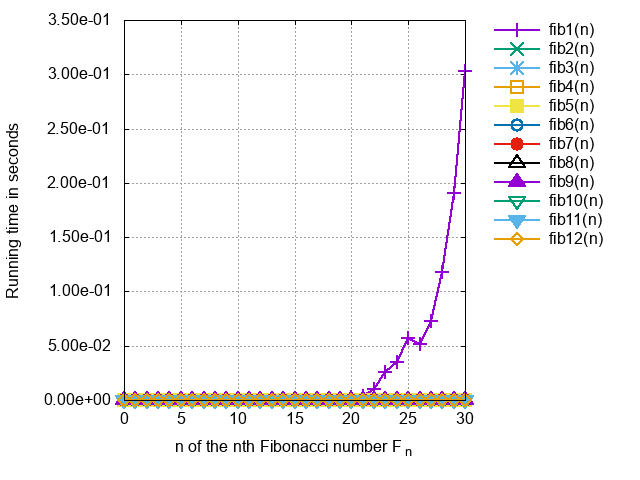}
  \caption{Results until \texttt{fib1} takes too much time.}
  \label{fig:fib_30}
\end{figure}

\begin{figure}
  \centering
  \includegraphics[width=0.7\textwidth]{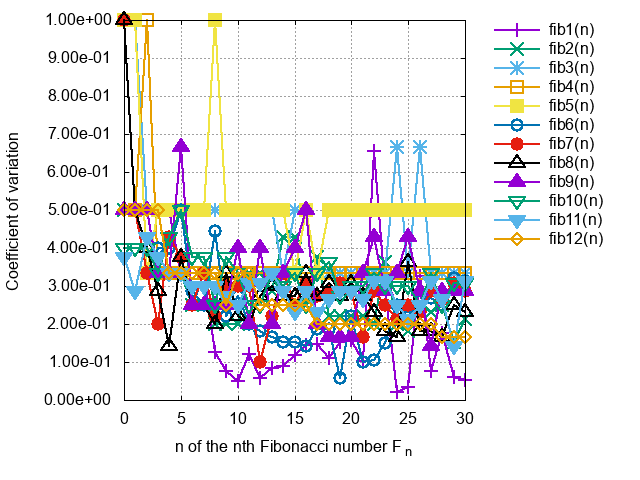}
  \caption{Coefficient of variation for above.}
  \label{fig:fib_30_cv}
\end{figure}

\subsection{Results with all algorithms}

The results with $0\leq n\leq 30$ in Fig.~\ref{fig:fib_30} mainly show
that the exponential-time algorithm \texttt{fib1} significantly
dominates all others in runtime. This algorithm is not practical at
all to use for larger $n$ to compute the Fibonacci numbers. The ratio
of the slowest runtime to the fastest runtime, that of \texttt{fib5},
is about four orders of magnitude, 14,000 to be exact on our
experimental setting.

The CV results in Fig.~\ref{fig:fib_30_cv} are not very reliable at
this scale, since the runtimes are so small. They still fall below
35\% for the most part, with an outlier at 50\% for \texttt{fib5}. For the
slowest algorithm \texttt{fib1}, the CV is largely around 5\%, which
is expected given its relatively large runtime.

\begin{figure}
  \centering
  \includegraphics[width=0.7\textwidth]{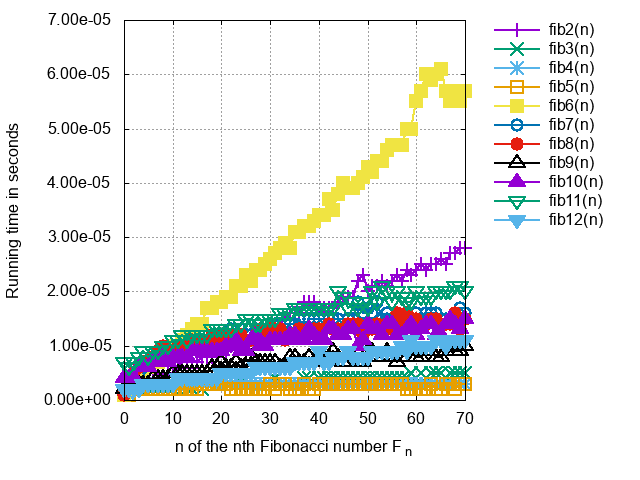}
  \caption{Results with all algorithms (excl. \texttt{fib1}) returning exact results.}
  \label{fig:fib_70}
\end{figure}

\begin{figure}
  \centering
  \includegraphics[width=0.7\textwidth]{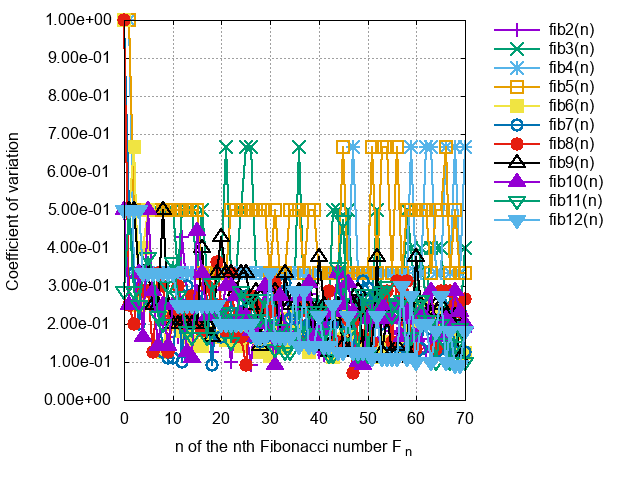}
  \caption{Coefficient of variation for above.}
  \label{fig:fib_70_cv}
\end{figure}

\subsection{Results with all fast algorithms returning exact results}

The results with $0\leq n\leq 70$ in Fig.~\ref{fig:fib_70} exclude the
slowest algorithm \texttt{fib1}. The results show that \texttt{fib6}
is now the slowest among the rest of the algorithms; its runtime grows
linearly with $n$, while the runtimes of the rest grow
sublinearly. The algorithms implementing the closed-form formulas run
in almost constant time.

The algorithms fall into the following bands of comparable runtime,
given in increasing order; within a band the differences are within
the measurement noise:
\begin{itemize}
\item \texttt{fib5}, \texttt{fib4}, \texttt{fib3};
\item \texttt{fib12}, \texttt{fib9};
\item \texttt{fib10}, \texttt{fib8}, \texttt{fib7}, \texttt{fib2}, \texttt{fib11};
\item \texttt{fib6}. 
\end{itemize}
These runtimes show that the algorithms implementing the closed-form
formulas are the fastest. The ratio of the slowest runtime, that of
\texttt{fib6}, to the fastest runtime, that of \texttt{fib5}, is about
an order of magnitude, 13 to be exact on our experimental setting.

The CV results in Fig.~\ref{fig:fib_70_cv} are similar to those of the
first case above.

\begin{figure}
  \centering
  \includegraphics[width=0.7\textwidth]{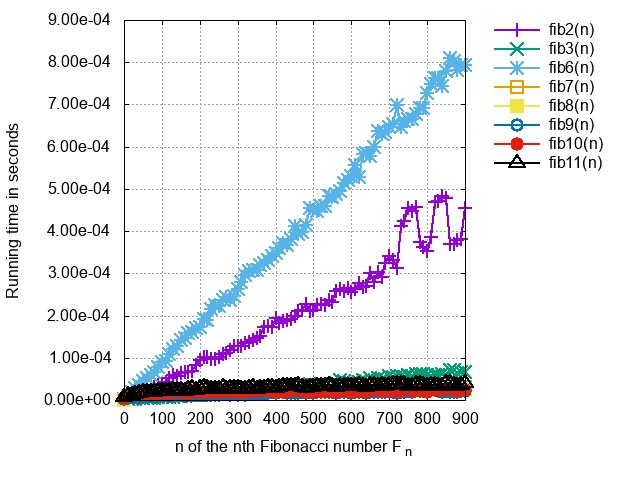}
  \caption{Results until recursive algorithms hit too deep a recursion depth.}
  \label{fig:fib_900}
\end{figure}

\begin{figure}
  \centering
  \includegraphics[width=0.7\textwidth]{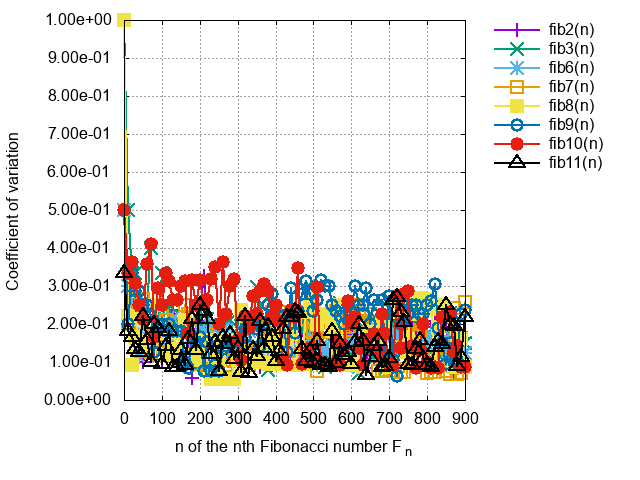}
  \caption{Coefficient of variation for above.}
  \label{fig:fib_900_cv}
\end{figure}

\begin{figure}
  \centering
  \includegraphics[width=0.7\textwidth]{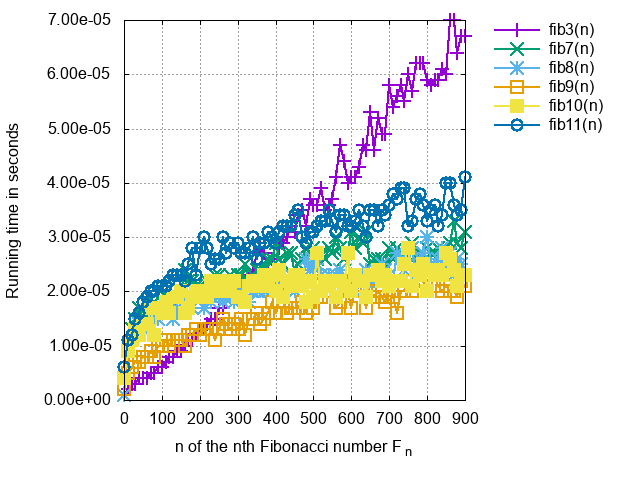}
  \caption{Results until recursive algorithms hit too deep a recursion
    depth. Faster algorithms focus.}
  \label{fig:fib_900_focus1}
\end{figure}

\begin{figure}
  \centering
  \includegraphics[width=0.7\textwidth]{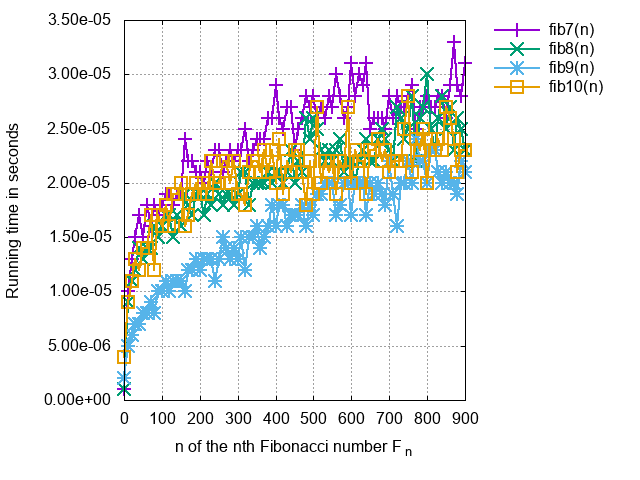}
  \caption{Results until recursive algorithms hit too deep a recursion
    depth. Even faster algorithms focus.}
  \label{fig:fib_900_focus2}
\end{figure}

\subsection{Results within safe recursion depth}

The results with $0\leq n\leq 900$ in Fig.~\ref{fig:fib_900} exclude
the slowest algorithm \texttt{fib1}. The upper bound of $n$ is roughly
the limit beyond which recursive algorithms fail due to the violation
of their maximum recursion depth.

The results show that \texttt{fib6} followed by \texttt{fib2} are the
slowest among the rest of the algorithms; their runtimes grow
linearly with $n$, though the slope for \texttt{fib2} is much
smaller. The remaining algorithms grow sublinearly. Again, the
algorithms implementing the closed-form formulas run in almost
constant time.

The algorithms group as follows in terms of their runtimes in
increasing runtime order:
\begin{itemize}
\item \texttt{fib9}, \texttt{fib10}, \texttt{fib8}, \texttt{fib7},
  \texttt{fib11}, \texttt{fib3};
\item \texttt{fib2};
\item \texttt{fib6}.
\end{itemize}
The ratio of the slowest runtime, that of \texttt{fib6}, to the
fastest runtime, that of \texttt{fib9}, is about two orders of
magnitude on our experimental setting.

Fig.~\ref{fig:fib_900_focus1} and Fig.~\ref{fig:fib_900_focus2} zoom
in on the faster algorithms, excluding the slowest algorithms from
Fig.~\ref{fig:fib_900}.

The CV results in Fig.~\ref{fig:fib_900_cv} are all below 20\%, which
is reasonably small given that the runtimes are now larger.

\begin{figure}
  \centering
  \includegraphics[width=0.7\textwidth]{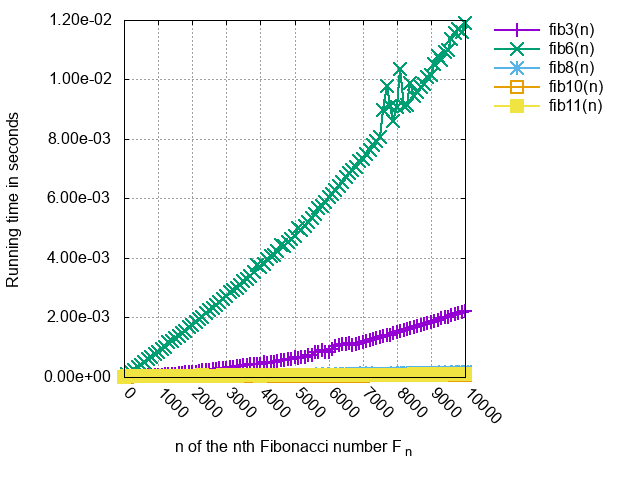}
  \caption{Results using only iterative algorithms with exact results.}
  \label{fig:fib_10k}
\end{figure}

\begin{figure}
  \centering
  \includegraphics[width=0.7\textwidth]{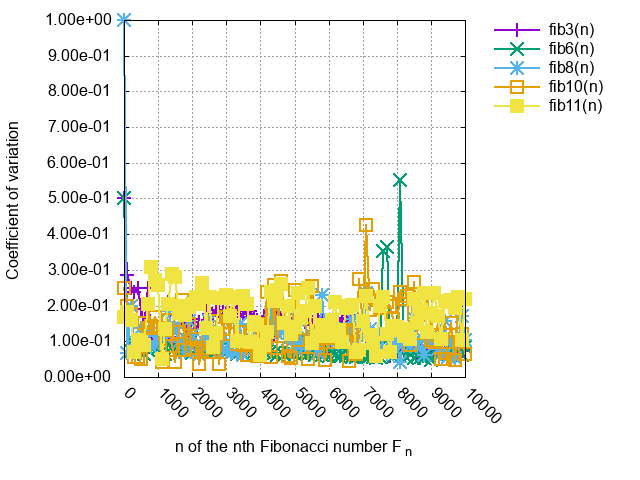}
  \caption{Coefficient of variation for above.}
  \label{fig:fib_10k_cv}
\end{figure}

\begin{figure}
  \centering
  \includegraphics[width=0.7\textwidth]{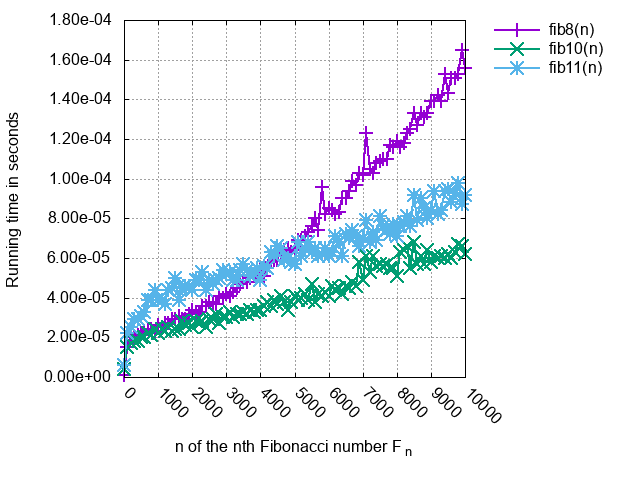}
  \caption{Results using only iterative algorithms with exact
    results. Faster algorithms focus.}
  \label{fig:fib_10k_focus1}
\end{figure}

\subsection{Results with the fastest iterative and exact algorithms}

The results with $0\leq n\leq 10{,}000$ in Fig.~\ref{fig:fib_10k} exclude
all slow algorithms, all recursive algorithms, and all inexact
algorithms, those that do not return exact results.

In this range where $n$ gets very large, the fastest and slowest
algorithms are \texttt{fib10} and \texttt{fib6}, respectively. The
algorithms group as follows in terms of their runtimes in increasing
runtime order:
\begin{itemize}
\item \texttt{fib10}, \texttt{fib11}, \texttt{fib8};
\item \texttt{fib3};
\item \texttt{fib6}.
\end{itemize}
The ratio of the slowest runtime, that of \texttt{fib6}, to the
fastest runtime, that of \texttt{fib10}, is about two orders of
magnitude, 130 to be exact on our experimental setting.

Fig.~\ref{fig:fib_10k_focus1} zooms in on the fastest algorithms,
excluding the slowest algorithms from Fig.~\ref{fig:fib_10k}.

The CV results in Fig.~\ref{fig:fib_10k_cv} have all converged to
values below 20\%, which is again reasonably small given how much
larger the runtimes are at this range of $n$.

\section{Conclusions}\label{sec:conclusions}

The Fibonacci numbers are well known and simple to understand. We have
selected from the technical
literature~\cite{CrStRi01,GrKnPa94,SoWe07,Wiki18} twelve methods of
computing these numbers. Some of these methods are recursive formulas
and some others are closed-form formulas. We have translated each
method into an algorithm and implemented it in the Python programming
language. Each algorithm takes in $n$ and returns the $n$th Fibonacci
number $F_n$.

Though simple, these algorithms illustrate a surprisingly large number
of concepts from the algorithms field, from top-down vs. bottom-up
dynamic programming through to recursion depth, and probably more.
Section~\ref{sec:intro} lists all of them, each with the algorithms
that illustrate it. The simplicity of these algorithms then becomes a
valuable asset in teaching introductory algorithms, in that students
can focus on the concepts rather than on the complexity of the
algorithms themselves.

We have also presented a small-scale experimental analysis of these
algorithms to further enhance their understanding. The analysis
reveals a couple of interesting observations, e.g., how two algorithms
that implement seemingly similar recursive formulas may have widely
different runtimes, how space usage can affect time, how or why
recursive algorithms cannot have too many recursive calls, when
approximate algorithms stop returning exact results, etc. The results
section explains these observations in detail with plots.

Probably the simplest fast algorithm to implement is
\texttt{fib3}, the one that uses constant space. However, the fastest
algorithm, especially for large $n$, turns out to be \texttt{fib10},
the one that implements a recursive algorithm with a logarithmic
number of iterations. When $n\leq 70$ where all algorithms return
exact results, the fastest algorithms are \texttt{fib4} and
\texttt{fib5}, the ones that implement the closed-form formulas.

The slowest algorithm is of course \texttt{fib1}, the one that
implements probably the most well-known recursive formula, which is
usually also the definition of the Fibonacci numbers. Memoization does
speed it up immensely but there is no need to add complexity when
simpler and faster algorithms also exist.

We hope that this paper can serve as a useful resource for students
learning and teachers teaching the basics of algorithms. All the
programs used for this study are at \cite{Da18}.

\section{Homework Questions}\label{sec:homework}

We now list a number of homework questions for the students. These
questions should help improve the students' understanding of the
algorithmic concepts further.

\begin{enumerate}
\item Try to simplify the algorithms further, if possible.
\item Try to optimize the algorithms further, if possible.
\item Replace $O$ with $\Theta$ in the complexity analyses, if possible.
\item Prove the time and space complexities for each algorithm.
\item Improve, if possible, the time and space complexities for each
  algorithm. A trivial way is to use the improved bounds on
  $M(\cdot)$.
  
\item Reimplement the algorithms in other programming languages. The
  simplicity of these algorithms should help in the speed of
  implementing in another programming language.
  
\item Derive an empirical expression for the runtime of each
  algorithm. This can help derive the constants hidden in the
  $O$-notation (specific to a particular compute setup).
\item Rank the algorithms in terms of runtime using statistically more
  robust methods than what this paper uses.
\item Find statistically more robust methods to measure the
  variability in the runtimes and/or to bound the runtimes.
\item Explain the reasons for observing different real runtimes for
  the algorithms that have the same asymptotic time complexity.
  
\item Learn more about the concepts related to recursion such as call
  stack, recursion depth, and tail recursion. What happens if there is
  no bound on the recursion depth?
\item Explain in detail how each algorithm illustrates the algorithmic
  concepts that Section~\ref{sec:intro} assigns to it, and argue for or
  against any assignment you disagree with.
  
\item Design and implement a recursive version of the algorithm
  \texttt{fib11}.
\item Design and implement versions of the algorithms \texttt{fib4}
  and \texttt{fib5} that use rational arithmetic rather than
  floating-point arithmetic. For the non-rational real numbers, use
  their best rational approximation to approximate them using rational
  numbers at different denominator magnitudes.
\item Find other formulas for the Fibonacci numbers and implement them
  as algorithms.
\item Prove the formula used by the algorithm \texttt{fib12} by induction
  on $n$, without appealing to the closed-form formula that
  Section~\ref{sec:algos} uses for it.

\item Use the formula involving binomial coefficients,
  $F_{n}=\sum_{k} \binom{n-k-1}{k}$, to compute the Fibonacci numbers.
  This will also help teach the many ways of computing binomial
  coefficients.
  
\item Design and implement multi-threaded or parallelized versions of
  each algorithm. Find out which algorithm is the easiest to
  parallelize and which algorithm runs faster when parallelized.
  
\item Measure, on your own machine, the largest $n$ for which
  \texttt{fib4}, \texttt{fib5} and \texttt{fib12} still return exact
  results. Why is the limit of \texttt{fib12} different from that of
  \texttt{fib4} and \texttt{fib5}? Of the limits reported in
  Section~\ref{sec:algos}, which would change under a different math
  library, and which follows from the IEEE 754 double precision format
  alone and so cannot?

\item Recompute the bit-operation column of Table~\ref{tab:cmp}, which
  assumes the schoolbook bound $M(b)=O(b^2)$, for Karatsuba
  multiplication, $M(b)=O(b^{1.585})$, and for an FFT-based
  multiplication. Find which algorithms change places, and whether the
  resulting ranking matches the measured runtimes for large $n$.

\item Explain why \texttt{fib9} and \texttt{fib10} differ in space
  complexity though they compute the same values from the same formula,
  then modify \texttt{fib9} so that it too uses $O(\lg n)$ space.

\item Explain what is missed by testing every algorithm in this paper
  against another algorithm from it, namely the one used by
  \texttt{fib3}, and design a test that catches it.

\item Repeat the measurements for negative $n$, which every algorithm
  accepts but the experiments never cover, and find whether the
  rankings change and whether they should.

\item Analyse the algorithms that the repository~\cite{Da18} has added
  beyond these twelve, place them in Table~\ref{tab:cmp}, and find
  which of the concepts of Section~\ref{sec:intro} each illustrates.

\item Responsibly update Wikipedia using the learnings from this study
  and/or your extensions of this study.
\end{enumerate}


\end{document}